\documentclass{aa}

\usepackage{graphicx}
\usepackage{txfonts}
\usepackage[colorinlistoftodos]{todonotes}
\usepackage{color}
\usepackage{booktabs}

\usepackage{amsmath}
\usepackage{natbib}
  \bibpunct{(}{)}{;}{a}{}{,}

\begin{document} 

\title{Dust in a compact, cold, high-velocity cloud: A new approach to removing  foreground emission}
\subtitle{}

\author{D. Lenz \inst{1,\thanks{\email{dlenz@astro.uni-bonn.de}}}, L. Fl\"oer\inst{1}, J. Kerp \inst{1}}
\institute{
Argelander-Institut f\"ur Astronomie (AIfA), Universit\"at Bonn, Auf dem H\"ugel 71, 53121 Bonn, Germany
}
\authorrunning{Lenz et al.}
\date{Accepted by A\&A December 13, 2015}

\abstract
	{Because isolated high-velocity clouds (HVCs) are found at great distances from the Galactic radiation field and because they have subsolar metallicities, there have been no detections of dust in these structures. A key problem in this search is the removal of foreground dust emission.}
	{Using the Effelsberg-Bonn \ion{H}{i} Survey and the Planck far-infrared data, we investigate a bright, cold, and clumpy HVC. This cloud apparently undergoes an interaction with the ambient medium and thus has great potential to form dust.}
	{To remove the local foreground dust emission we used a regularised, generalised linear model and we show the advantages of this approach with respect to other methods. To estimate the dust emissivity of the HVC, we set up a simple Bayesian model with mildly informative priors to perform the line fit instead of an ordinary linear least-squares approach.}
	{We find that the foreground can be modelled accurately and robustly with our approach and is limited mostly by the cosmic infrared background. Despite this improvement, we did not detect any significant dust emission from this promising HVC. The $3\sigma$-equivalent upper limit to the dust emissivity is an order of magnitude below the typical values for the Galactic interstellar medium.}
    {}


\maketitle


\section{Introduction}
\label{ch:intro}

Since their discovery by \citet{muller1963}, high-velocity clouds (HVCs) have been the target of a wide range of studies (see \citet{wakker1997} for a review). It is thought that they are located at distances of several kpc \citep{wakker2001} and contribute to the fuelling of low-metallicity gas into the Galaxy \citep{putman2012}.

Early attempts to detect dust emission from HVCs were unsuccessful \citep{wakker1986, desert1988} and were generally considered to be difficult because of the clouds' low metallicities and hence low dust-to-gas ratios \citep{fox2004}. Moreover, HVCs are located far from the interstellar radiation field (ISRF), which implies a faint illumination by UV light that is absorbed and re-emitted by the dust grains. The cosmic infrared background radiation (CIB) has anisotropies on angular scales that are comparable to the typical sizes of HVCs and is therefore another source of confusion \citep{planck2011_xxiv, planck2014_xxx}. Very recently, the \citet{planck2014_xvii} has reported that the variation of dust emissivities across the field of interest is the limiting source of uncertainty when modelling the dust data.

Recent attempts to disclose the far-infrared (FIR) emissivity of HVCs are in line with these findings. Neither the investigation of different high-latitude clouds \citep{planck2011_xxiv} nor the stacking of GALFA-\ion{H}{i} compact clouds \citep{saul2014} has detected significant FIR emission.

Despite these odds, \citet{miville2005b} generate a simple model of the dust emission from HVC complex C based on its \ion{H}{i} column density and find a faint, but significant dust emissivity for the HVC regime. A similar approach, combined with an investigation of the chance correlation of \ion{H}{i} and dust emission, yields a $>3\sigma$ detection of dust towards complex M \citep{peek2009}.

Here we investigate the HVC located at $(l,b,v_{\rm LSR})$ = $(125\,^{\circ}, 41\,^{\circ}, -207\,\rm km\,s^{-1}$), hereafter HVC125. The cloud has been previously studied with the Effelsberg 100 m telescope \citep{bruens2001} and the Westerbork Synthesis Radio Telescope \citep{braun2000}. In combination, these data sets disclose a two-phase structure and a head-tail morphology of the HVC of interest. This might indicate ram-pressure interaction which in turn results in a reduced formation time of H$_2$ and dust \citep{guillard2009, roehser2014}. The single-dish data suggest that the warm phase is stripped off the core of the HVC. The cold component has an extraordinarily narrow line width around $\Delta v = 2\,\rm km\,s^{-1}$ FWHM, equivalent to a kinetic temperature of $T_{\rm kin} \lesssim 85\,\rm K$. The compact spatial structure of a few arcminutes, high brightness temperature of $T_B \gtrsim 10\,\rm K$ in the single-dish data, and low kinetic temperature make HVC125 one of the most promising HVCs in terms of detection probability of FIR dust emission.

\section{Data}
\label{ch:data}

We used data from the recently finished Effelsberg-Bonn \ion{H}{i} Survey \citep[EBHIS,][]{kerp2011, winkel2010, winkel2016} to study the neutral atomic hydrogen in HVC125.

For the FIR data, we used the latest release of the Planck data at $857\,\rm GHz$ \citep{planck2015_viii}. We chose the highest frequency of the Planck data because the relative contributions of the cosmic microwave background and the cosmic infrared background to the uncertainty in modelling the foreground both decrease in proportion to the frequency \citep[][their Fig. C.1]{planck2014_xvii}.

To account for the differences in angular resolution, the dust data were smoothed to the angular resolution of the EBHIS data by Gaussian convolution. The final angular resolution is $10.83'$. The EBHIS data have a spectral resolution of $1.49\,\rm km\,s^{-1}$ at a channel spacing of $1.28\,\rm km\,s^{-1}$.

\section{Analysis}
\label{ch:analysis}

In the following, we use $N$ to refer to the \ion{H}{i} column density $N_{\ion{H}{i}}$ and $I$ to refer to the FIR intensity at $857\,\rm GHz$. When analysing the dust content of HVCs by comparing their \ion{H}{i} column density to their dust content, the most challenging step is the estimation of the foreground dust emission. Because of the complexity and uncertainty of the correlation of dust and gas, an accurate and robust determination of the foreground component is of the utmost importance \citep[e.g.][]{peek2009, saul2014}.

\begin{figure*}[tp]
    \includegraphics{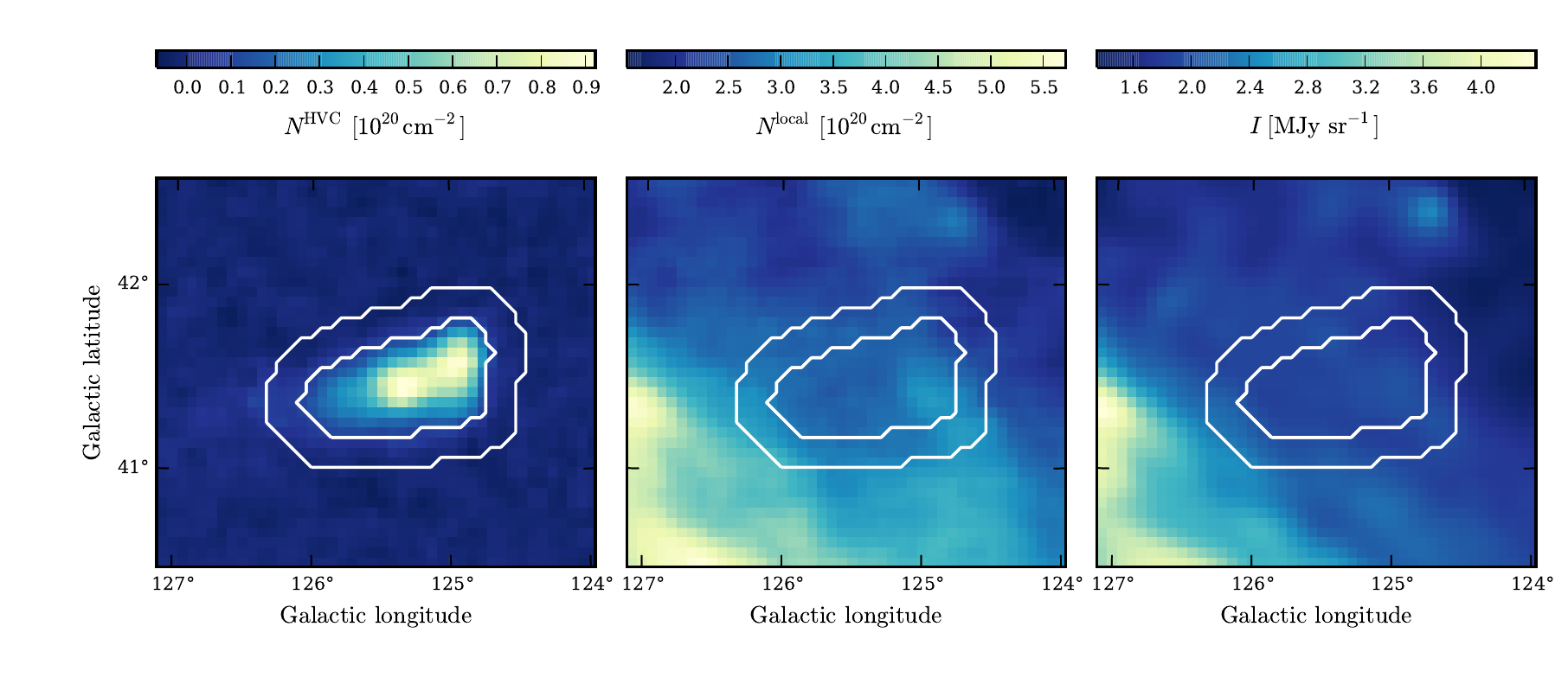}
	\caption{\textbf{Left:} EBHIS column density map $N^{\rm HVC}$ of the HVC ($-230\,\mathrm{km\,s^{-1}} < v_{\rm LSR} < -190\,\mathrm{km\,s^{-1}}$) \textbf{Center:} EBHIS column density map $N^{\rm local}$ of the local foreground emission ($-190\,\mathrm{km\,s^{-1}} < v_{\rm LSR} < +30\,\mathrm{km\,s^{-1}}$). \textbf{Right:} FIR intensity $I_{\rm 857\,GHz}$ from \citet{planck2015_viii}. The inner white contour line corresponds to the $5\sigma$ noise level in the HVC column density map. The outer contour marks an annulus that contains the same number of pixels as the tight HVC mask.}
	\label{fig:coldenses}
\end{figure*}

The \ion{H}{i} data allow us to distinguish between local foreground emission and HVC emission via the radial velocity. For HVC125, we selected the velocity range $(-230\,\mathrm{km\,s^{-1}}, -190\,\mathrm{km\,s^{-1}})$ for the HVC and the remaining range $(-190\,\mathrm{km\,s^{-1}}, +30\,\mathrm{km\,s^{-1}})$ for the foreground emission. The corresponding column density maps are shown in Fig. \ref{fig:coldenses}. Moreover, we present an image of the FIR intensity at $857\,\rm GHz$ in the direction of HVC125. The inner contour outlines the  $5\sigma$ level of the HVC \ion{H}{i} column density. The outer contour marks an annulus around the HVC that contains as many pixels as the inner, narrow mask. In the following, we use these masks to determine the foreground dust contribution and the dust emissivity of the HVC.

For this analysis, we decided not to include the statistical uncertainties from the data noise of $\sigma_{\rm RMS} = 90\,\rm mK$ for the EBHIS data and $\sigma_{\rm RMS} = 0.014\,\rm MJy\,sr^{-1}$ for the Planck data at $857\,\rm GHz$. We refer again to Fig. C.1 in \citet[][]{planck2014_xvii}, which shows that at this frequency the analysis is dominated by uncertainties from the foreground estimation.

\subsection{Standard approach}
\label{sect:method_standard}

The standard approach \citep[e.g.][]{miville2005b, planck2011_xxiv} used to evaluate  the dust content of HVCs is the superposition of different \ion{H}{i} column density maps $N^i$ to model the FIR intensity:
\begin{equation}
    I(x,y) = \epsilon^{\rm local} N^{\rm local}(x,y) + \epsilon^{\rm HVC} N^{\rm HVC}(x,y) + Z
    \label{eq:classic}
\end{equation}
Here, $I$ is the FIR intensity, $\epsilon$ denotes the dust emissivity per hydrogen nucleon, and $Z$ is a constant offset; $I$ and $N^i$ are two-dimensional images with $(x,y)$ denoting the spatial position of the pixel.

We fitted Eq. (\ref{eq:classic}) with a least-squares approach to quantify the parameters $\epsilon^{\rm local}$, $\epsilon^{\rm HVC}$, and $Z$. To investigate the influence of the spatial area that is fitted on the results, we performed the fit for three different spatial masks: a tight mask around the HVC, a more extended mask, and the full image. The results are compiled in Table \ref{tab:classic}. The large variations for different spatial masks and the uncertainties on the fit parameters emphasise that this approach is very sensitive to changes in the area of interest. Furthermore, this approach only relies on two parameters (Eq. \ref{eq:classic}) to describe the foreground: the local dust emissivity $\epsilon^{\rm local}$ and the offset $Z$ and so it cannot account for multiple features at different radial velocities, possibly exposed to different physical environments. For the full field, we show the modelled FIR intensity $I^{\rm Standard}$ and the residual $I - I^{\rm Standard}$ in Fig. \ref{fig:4panel}, left column.

\begin{table}[tp]
\centering
\begin{tabular}{c|c|c|c}
    Mask & $\epsilon^{\rm local}$ & $\epsilon^{\rm HVC}$ & $Z$\\
    \hline
    \hline
    Small & $0.26\pm 0.48$ & $0.05\pm 0.38$ & $1.14\pm 1.31$\\
    Wide & $0.41 \pm 0.09$ & $0.029\pm 0.11$ & $0.76\pm 0.26$\\
    None & $0.61\pm 0.005$ & $-0.05\pm 0.03$ & $0.26\pm 0.02$\\
\end{tabular}
\caption{Emissivities $\epsilon^{i}$ for the different velocity components and offset $Z$, according to Eq. (\ref{eq:classic}). Units are $\rm MJy\,sr^{-1}/10^{20}\,\rm cm^{-2}$ for the emissivities and $\rm MJy\,sr^{-1}$ for the offset. The fit uncertainties are asymptotic standard errors, taken from the covariance matrix.}
\label{tab:classic}
\end{table}

\subsection{Generalised linear model for foreground estimation}
\label{sect:glm}

To overcome the limitations of the standard approach, we applied a generalised linear model \citep[GLM; for a review see][]{madsen2010, desouza2014} to the data. For this, we assume that each channel $T_B^i$ of the \ion{H}{i} cube can contribute individually to the FIR intensity. In consequence, we do not rely on a vague definition of the velocity range for local and HVC gas.

Within the GLM, the FIR intensity can be written as 
\begin{equation}
    I^{\rm local}(x,y) =  \sum_i T_B^i(x,y) \beta^i + Z.
    \label{eq:glm}
\end{equation}
The $\beta^i$ are the GLM coefficients and can be understood as emissivity per spectral channel. The parameter $Z$ is a global offset to the model. Because of the co-linearity between neighbouring spectral channels, the assumption of independent data for ordinary least-squares fitting is violated. To break this degeneracy, we controlled the GLM with lasso regularisation \citep{tibshirani1996} and minimised the term
\begin{equation}
    ||I^{\rm local}-T_B^i\beta^i-Z||^2 + \alpha\cdot|| \beta^i||_1 .
    \label{eq:elastic_net}
\end{equation}
The first part of this term corresponds to the regular least-squares approach and gives the residual sum of squares. The second part is the penalty term and ensures that the coefficients $\beta^i$ are chosen as sparsely as possible. The strength of this second term is scaled by $\alpha$. We used simulations and cross validation to optimise this regularisation strength (see Sect. \ref{sect:glm_verification}). The regularised GLM and the cross validation was implemented via \texttt{scikit-learn} \citep{pedregosa2011}, a machine-learning package for \texttt{python}.

For our application to the \ion{H}{i} and dust data, only data points outside of the wide mask and with $v_{\rm LSR} > -190\,\rm km\,s^{-1}$ were considered because they are unrelated to the HVC. This ensures that the HVC signal is not accidentally removed by our foreground subtraction. While the distinction between HVC and local gas is very straightforward in this particular case, there are other cases where gas at high or intermediate velocities is difficult to disentangle \citep[e.g. HVC Complex M,][]{wakker2001}. In these cases, we cannot simply remove the foreground by applying a threshold in radial velocity, but have to rely only on spatial masking of the HVC.

\begin{figure}[tp]
    \includegraphics{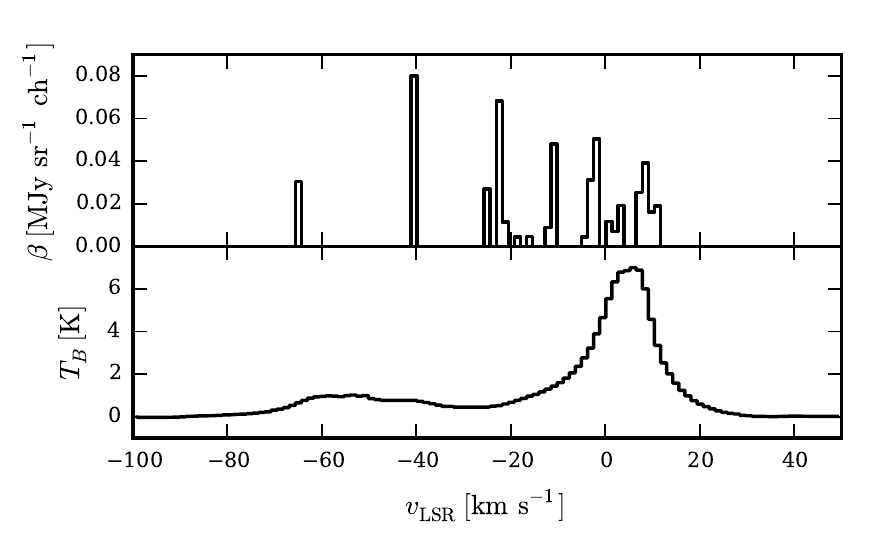}
    \caption{\textbf{Top:} GLM coefficients $\beta^i$ for each channel based on the cross-validated lasso regression (Eqs. \ref{eq:glm} and \ref{eq:elastic_net}). \textbf{Bottom:} Mean \ion{H}{i} spectrum of the data cube.}
    \label{fig:en_coeffs_spectrum}
\end{figure}

The resulting GLM coefficients $\beta^i$ and the mean \ion{H}{i} spectrum are shown in Fig. \ref{fig:en_coeffs_spectrum}. The offset is $Z = 0.73\,\rm MJy\,sr^{-1}$. We find that the dust emission towards HVC125 can be modelled well with approximately seven different emissivities. The narrowness of the GLM coefficients is the product of the applied regularisation. Visually inspecting the HI data cube discloses that each individual GLM coefficient is indeed associated physically with an individual HI structure. However, the exact matching of \ion{H}{i} structures to FIR emissivity peaks varies with the regularisation strength and demands further justification (Sect. \ref{sect:glm_verification}). The sparse filling of the whole HI spectrum with GLM coefficients argues strongly for a localisation of the dust within the cold neutral medium (CNM) filaments and is inconsistent with a homogenous mixture of dust and gas on all linear scales. However, we note that the discrete nature of the FIR emissivities for each spectral channel is primarily a schematic description.

\begin{figure}[tp]
    \includegraphics{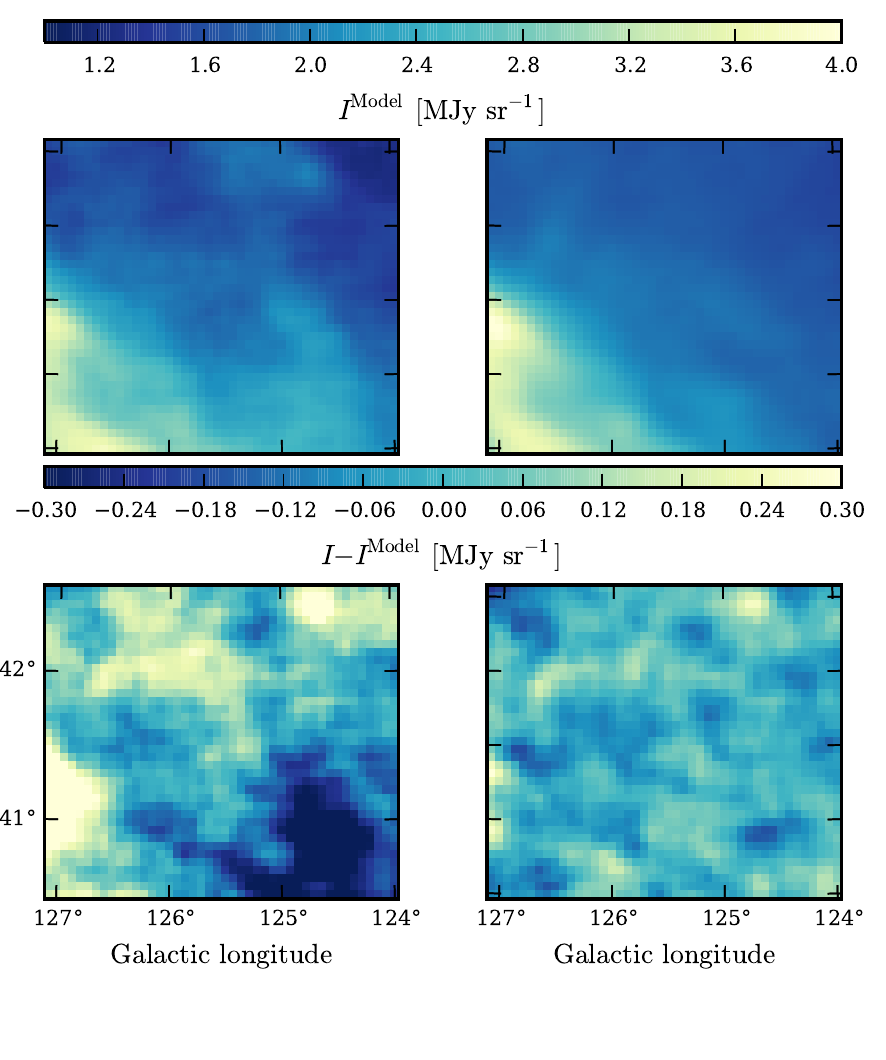}
    \caption{Model of the FIR intensity (top) and residual emission (bottom). We show the results of the standard approach (Sect. \ref{sect:method_standard}) in the left column and the GLM (Sect. \ref{sect:glm}) in the right column.}
    \label{fig:4panel}
\end{figure}

Similar to the standard approach, we show the modelled FIR intensity and the residual (Fig. \ref{fig:4panel}, right column). A comparison of the two different approaches shows that the GLM can cover a wider range of FIR intensities and manages to account for a larger number of  different features. The residual is less structured and the remaining structures stem from the CIB anisotropies (compare Sect. \ref{sect:glm_verification}).

\begin{figure}[tp]
    \includegraphics{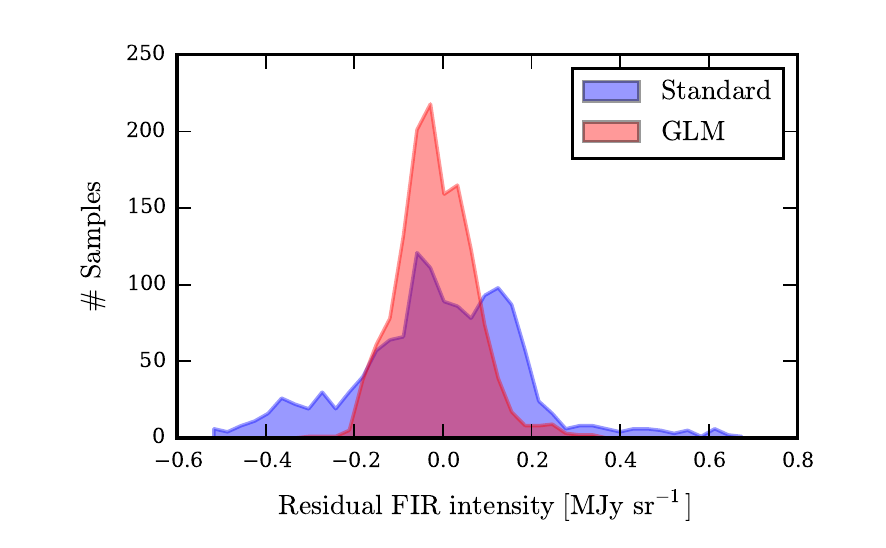}
    \caption{Residual histograms after correcting for the local foreground FIR intensity outside of the wide mask for the two different methods.}
    \label{fig:residual_histograms}
\end{figure}

Furthermore, we show the histograms from the residual FIR intensity in Fig. \ref{fig:residual_histograms}. For both models, we consider only data points outside of the wide mask because we are only interested in the capability to model the foreground emission. For the standard approach, we find that the residual is rather broad and irregular. In contrast, the GLM residual is narrow and symmetric and its shape can be approximated as Gaussian.

\subsection{Measurement of the HVC dust content}
\label{sect:hvc_dust}

To quantify the hypothetical dust emission from HVC125, we corrected the FIR intensity map for the local foreground emission by using the GLM, $I^{\rm HVC} = I - I^{\rm GLM}$ (Fig. \ref{fig:4panel}, bottom right). We investigated the correlation between the HVC \ion{H}{i} column density and the HVC FIR intensity within the narrow mask with a linear correlation in the Bayesian framework \citep[e.g.][]{dagostini2013}. Thus, we sampled the posterior given by
\begin{equation}
    p(\epsilon, \sigma|\mathcal{D}) \propto \mathcal{L} (\mathcal{D}|\epsilon, \sigma)p(\epsilon, \sigma)
    \label{eq:posterior}
\end{equation}
where $\mathcal{D}$ is the data vector $\mathcal{D} = (N^{\rm HVC}, I^{\rm HVC})$. The likelihood is given by
\begin{equation}
    \mathcal{D} | \epsilon, \sigma \sim \mathcal{N}(I^{\rm HVC}- \epsilon\cdot N^{\rm HVC},\sigma^2).
    \label{eq:likelihood}
\end{equation}
The parameter $\epsilon$ is the dust emissivity per hydrogen nucleus and $\sigma$ is the standard deviation of the Gaussian likelihood.

We selected minimally informative priors for our parameters:
\begin{align}
    \label{eq:prior}
    \epsilon & = \tan\phi\nonumber\\
    p(\phi) & = \mathrm{Uniform}(\phi, 0, \pi/2)\\
    p(\log\sigma) & = \mathrm{Uniform}(\log\sigma, -8, 4)\nonumber
\end{align}
For the emissivity $\epsilon$, we sampled uniformly in $\arctan(\epsilon)$ to avoid a bias towards larger values. A simple scale-invariant prior was chosen for the scatter $\sigma$ \citep{jeffreys1946} in a reasonable range. Furthermore, we reasonably chose to restrict $\epsilon$ to positive values. An offset FIR intensity is already part of the GLM, hence we choose not to include a further offset here. Lastly, we note that we did not account for the spatial covariance in the image owing to the non-flat CIB angular power spectrum. For a complete and proper treatment of this effect, we would also require an accurate determination of the spatial covariance due to the beam shapes of the different data sets and the sampling on the pixel grid.

\begin{figure}[tp]
    \includegraphics{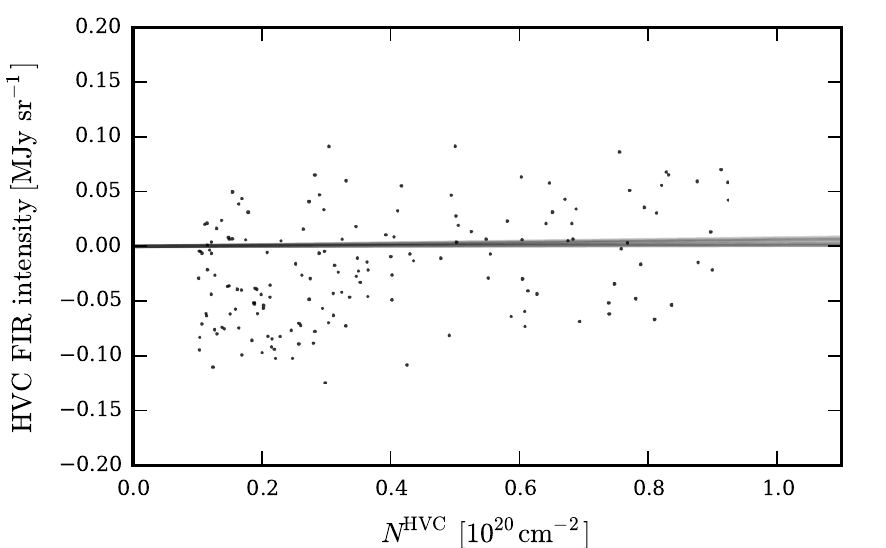}
    \caption{Linear correlation between HVC \ion{H}{i} column density and foreground-subtracted FIR intensity in the narrow mask. The lines correspond to thirty randomly chosen MC samples (Eqs. \ref{eq:posterior} to \ref{eq:prior}) from the Bayesian line fit.}
    \label{fig:fit}
\end{figure}

\begin{figure}[tp]
    \includegraphics{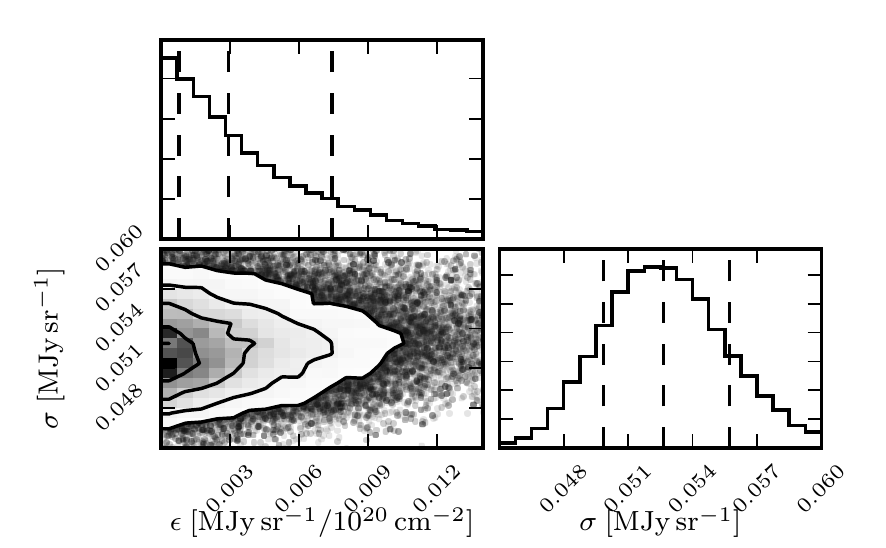}
    \caption{Posterior distribution for $\epsilon$ and $\sigma$, obtained from sampling Eq. (\ref{eq:posterior}). The lines in the histogram indicate the 16th, 50th, and 84th percentile, equivalent to mean and $\pm 1\sigma$ for a Gaussian posterior.}
    \label{fig:triangle}
\end{figure}

The model was sampled with \texttt{emcee} \citep{foreman2013}, a \texttt{python} implementation of the affine-invariant ensemble sampler for Markov chain Monte Carlo \citep[MCMC,][]{goodman2010}. Figure \ref{fig:fit} shows randomly drawn samples of this model, applied to the narrow mask \ion{H}{i} and dust data of HVC125. The posterior distribution of the individual parameters is presented in Fig. \ref{fig:triangle}.

As in the standard approach (Table \ref{tab:classic}), we not only evaluated the narrow mask, but also the wide mask and the full field. The resulting fit parameters are summarised in Table \ref{tab:modern}.

We find that the emissivity $\epsilon$ and scatter $\sigma$ parameters are well-sampled. The HVC emissivity is not normally distributed and illustrates that the model strongly prefers zero emissivity. The 99.87\,\% upper limit, corresponding to $3\sigma$, is 0.021$\,\rm MJy\,sr^{-1}/10^{20}\,\rm cm^{-2}$ and thus an order of magnitude below typical Galactic ISM values \citep{planck2011_xxiv}. Moreover, the parameters vary only slightly for different masks. As expected, the scatter increases with the size of the mask, or equivalently with the number of data points.


\begin{table}[tp]
\centering
\begin{tabular}{c|ccc|ccc}
    \toprule
    \toprule
    & \multicolumn{3}{c|}{$\epsilon\ [\rm MJy\,sr^{-1}/10^{20}\,\rm cm^{-2}]$} & \multicolumn{3}{c}{$\sigma\ [\rm MJy\,sr^{-1}]$}\\
    Mask & 16\% & Median & 84\% & 16\% & Median & 84\%\\
    \midrule
    Small & 0.001 & 0.003 & 0.007 & 0.049 & 0.053 & 0.056\\
    Wide & 0.001 & 0.002 & 0.006 & 0.045 & 0.047 & 0.049\\
    None & 0.001 & 0.003 & 0.008 & 0.064 & 0.065 & 0.066\\
\end{tabular}
\caption{Emissivity $\epsilon$ and scatter $\sigma$ for the different masks. The numbers indicate the median of the posterior distribution and the 16th and 84th percentile. For a Gaussian posterior, this is equivalent to median $\pm 1\sigma$.}
\label{tab:modern}
\end{table}

\section{Verification of the GLM}
\label{sect:glm_verification}

In the following, we present our thorough investigation of the performance of the GLM; we conducted a series of simulations and tests to explore its advantages and limitations. Moreover, we used these simulations to properly select the regularisation strength $\alpha$ (Eq. \ref{eq:elastic_net}).

\subsection{Construction of the simulations}

The simulations were generated in the following way:
\begin{enumerate}
    \item We generated an artificial, noisy spectrum of GLM coefficients;
    \item we convolved this coefficient spectrum with the measured \ion{H}{i} data cube to generate a map of foreground FIR intensity;
    \item We added a random realisation of the CIB to the foreground.
\end{enumerate}

\begin{figure}[tp]
    \includegraphics{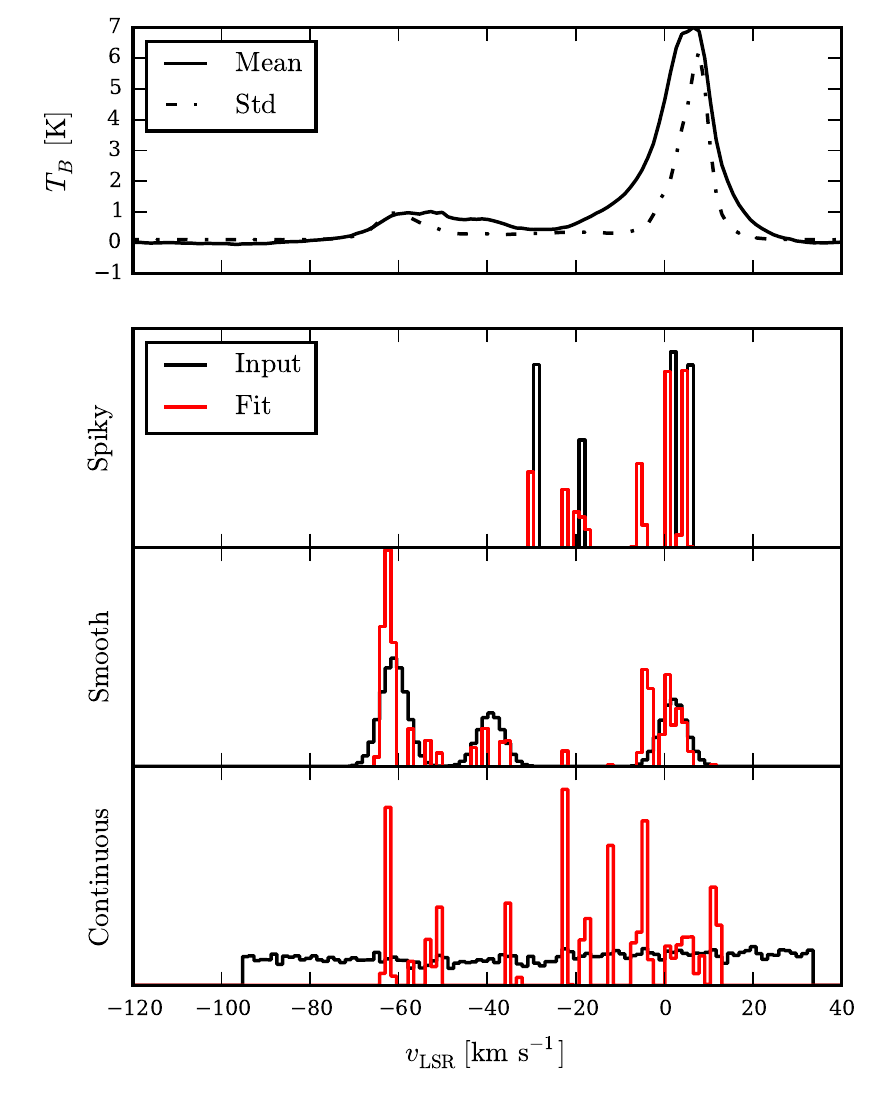}
    \caption{Top panel: Spectrum of mean and standard deviation of the full \ion{H}{i} data cube. Bottom panels: Input spectra (black) of GLM coefficients to simulate FIR intensity maps. The reconstruction by the GLM is shown in red and has been shifted to the left by one channel for illustration purposes. See the text for a detailed description.}
    \label{fig:simulation_input}
\end{figure}

Because we did not know the real nature of dust emissivities at different radial velocities, we tested different approaches. For all of them, we assumed that the dust emissivity occurs only in spectral channels in which significant \ion{H}{i} emission is found. We distinguished between the following types of GLM coefficients (see also Fig. \ref{fig:simulation_input}):

\begin{description}
    \item [\textbf{Spiky:}] One spectral channel wide, between 2 and 7 components. The spikes vary in amplitude by up to 40\,\%.
    \item [\textbf{Smooth:}] Similar to the spiky input, but the spectrum of GLM coefficients is smoothed with random Gaussian kernels with FWHM between $2\,\rm km\,s^{-1}$ and $8\,\rm km\,s^{-1}$.
    \item [\textbf{Flat:}] This represents the standard approach to investigating the \ion{H}{i}-FIR correlation. We choose a constant dust emissivity with 10\,\% random fluctuation for local gas ($v_{\rm LVC} > -30\,\rm km\,s^{-1}$) and a 20\,\% lower dust emissivity for the gas at intermediate velocities. ($-95\,\rm km\,s^{-1} < v_{\rm IVC} < -30\,\rm km\,s^{-1}$)
\end{description}

The amplitudes of all these GLM coefficient spectra were normalised such that the resulting mean intensity of the foreground FIR map was equal to the value found in the measured FIR intensity map (Fig. \ref{fig:coldenses}, right panel). This ensured that we modelled the proper ratio of foreground to background.

To combine the foreground FIR intensity with the CIB, we generated random realisations of the Gaussian random field based on the CIB angular power spectrum taken from \citet{planck2014_xxx}. We extrapolated the angular power spectrum from their Table D.2 with a power-law. This extrapolation does not hold for large angular separations i.e. small multipoles, but this effect is negligible for the present field size of only $2^{\circ}\times 3^{\circ}$. Here,  a power law is a valid approximation (G. Lagache, priv. comm.). To obtain the proper ratio of foreground and background components, we also scaled the simulated CIB to match the mean and fluctuation amplitude given in \citet[their Table 5]{planck2011_xviii}. Despite the non-flat angular power spectrum of the CIB, this is possible because the angular size of the fields investigated in \citet{planck2011_xviii} is similar to the field size in the present study. Finally, the simulated CIB was smoothed to the angular resolution of $10.8'$, which was used for all data throughout this study.

We generated 1000 of these simulations for each type of spectrum (spiky, smooth and flat) and applied the GLM to reconstruct the total FIR intensity. We compare the input and outcome of the GLM coefficient spectra in Fig. \ref{fig:simulation_input}.

\subsection{GLM performance on simulations and choice of regularisation strength}
\label{sect:simulation_evaluation}

The visual inspection of Fig. \ref{fig:simulation_input} shows that, despite the CIB confusion, we were able to properly reconstruct the shape and position of the GLM coefficients for the spiky and smooth case. Because of the GLM design, we did not recover the exact shape of the flat input spectrum, but used the most relevant channels to create an approximation that produces an accurate model. We note again that the discrete description of the dust emissivities for the spectral channels is a result of our approach and does not necessarily reflect the physical conditions.

The results are furthermore evaluated via three quantities: the mean strength of the CIB, the strength of its fluctuations and the Pearson's $r$ correlation coefficient of the reconstructed and the input CIB image. This mainly ensures that the GLM neither over- nor underfits the data. For the CIB mean $\mu$ and fluctuation amplitude $\sigma$, the values are taken from \citet{planck2011_xviii}.

\begin{figure}[tp]
    \includegraphics{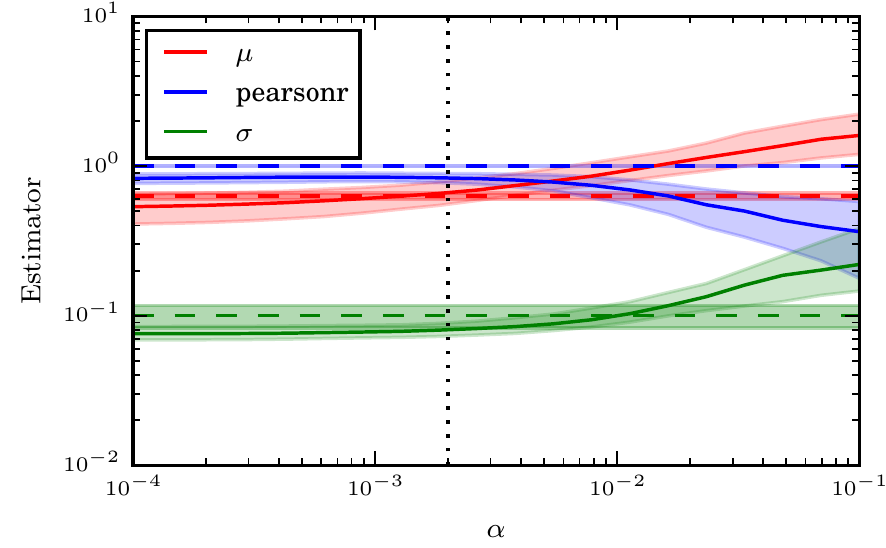}
    \caption{Evaluation of the reconstructed CIB mean $\mu$, CIB fluctuation amplitude $\sigma$ and Pearson's $r$ of input CIB image and reconstruction. This is based on the smooth input GLM coefficients (third panel from the top in Fig. \ref{fig:simulation_input}). Dashed lines indicate the input, solid lines the reconstructed quantities. Contours correspond to $1\sigma$ uncertainties. The vertical line indicates our choice of $\alpha$.}
    \label{fig:simulation_evaluation}
\end{figure}

We investigated how these quantities vary as a function of the regularisation strength $\alpha$ (Fig. \ref{fig:simulation_evaluation}). Our example was generated for the smooth input GLM coefficients. For the other cases, the results are presented in Sect. \ref{ch:appendix_estimator}. Dashed lines indicate the input values, solid lines are the results of our application of the GLM to the simulated data. The contours correspond to $1\sigma$ uncertainties.

We find that for $\alpha \lesssim 4\times 10^{-3}$, there is an agreement between input and reconstruction for all three estimators within their respective uncertainties. We chose to set $\alpha = 2\times 10^{-3}$ for our analysis of HVC125 (vertical line in Fig. \ref{fig:simulation_evaluation}).

The \textbf{Pearson's $r$} correlation coefficient between the input and reconstructed CIB is remarkably constant and close to 1, even for a very weak regularisation that allows a great number of GLM coefficients. This illustrates that chance correlation by individual \ion{H}{i} spectral channels is not very efficient in mimicking the CIB signal. For a very strong regularisation, the model cannot account for all the dust-emitting \ion{H}{i} components and the residual map is dominated by foreground emission, yielding a poor correlation to the input CIB.

The \textbf{CIB mean $\mu$} is the quantity that varies strongest for different values of the regularisation strength $\alpha$. We find that for $\alpha = 10^{-3}$ to $\alpha = 4\times 10^{-3}$, the CIB mean is properly estimated.

The \textbf{CIB fluctuation amplitude $\sigma$} is systematically underestimated if the regularisation is too weak, meaning that the model overfits the data. Other than in the Pearson's $r$ correlation, this hints towards a mimicking of background CIB by chance correlation with some \ion{H}{i} channels. This effect, however, is of the order of $10\,\%$ and is within the uncertainties for our choice of $\alpha = 2\times 10^{-3}$.

Furthermore, we find that the choice of $\alpha$ and the evaluation of the different quality estimators does not vary strongly for different types of input GLM coefficients (Fig. \ref{fig:simulation_evaluation_spiky}, \ref{fig:simulation_evaluation_flat}). Because we cover a variety of shapes (Fig. \ref{fig:simulation_input}) and demonstrate that the GLM can properly remove the foreground and uncover the faint CIB emission in each individual case, we conclude that it is well suited for the search of dust in HVCs.

\begin{figure}[tp]
    \includegraphics[bb=0 35 400 240, clip=]{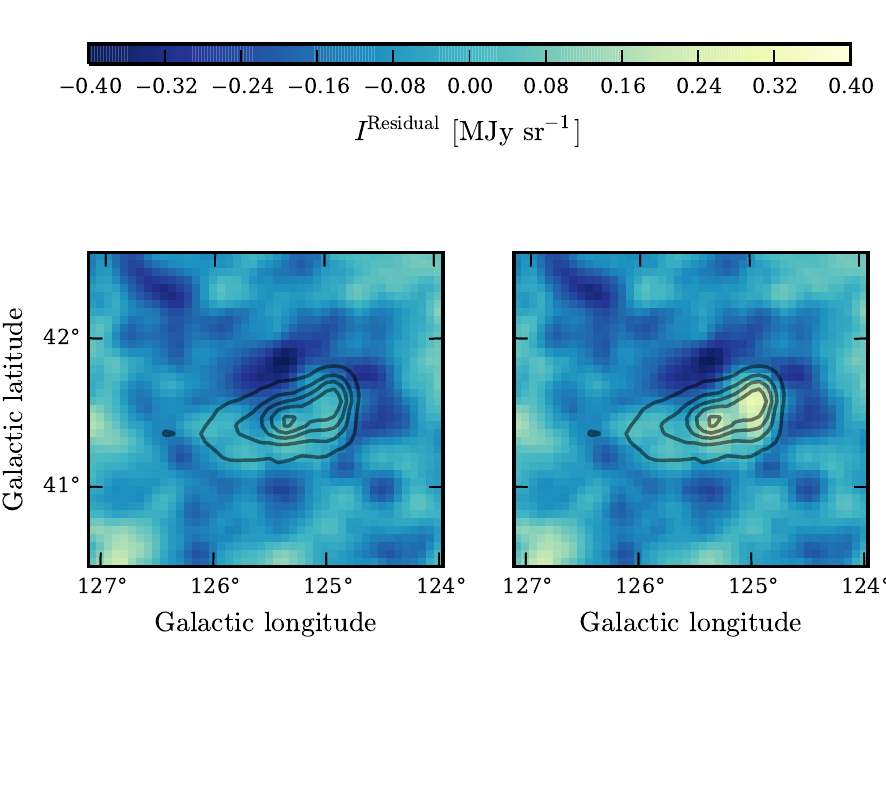}
    \caption{GLM reconstruction of a simulated dusty HVC. The image is dominated by the simulated CIB anisotropies. The HVC dust emissivity corresponds to $10\%$ (left) and $30\%$ (right) of typical Galactic values. The black contours correspond to the HVC column density starting at $\rm 10^{19}\,cm^{-2}$ and increasing in steps of $\rm 2\times 10^{19}\,cm^{-2}$.}
    \label{fig:dusty_hvc}
\end{figure}

To conclude, our approach is not constructed to precisely measure the dust emissivity of the individual clouds and filaments; rather, it is designed to remove the local foreground for studies of faint FIR signals such as CIB emission or dust in HVCs. To further verify this, we simulated a FIR intensity map in which the HVC has a dust emissivity of $10\%$ and $30\%$ of typical Galactic values and removed the foreground emission (Fig. \ref{fig:dusty_hvc}). We find that for an emissivity of only $10\%$ Galactic, the HVC can barely be distinguished from the residual CIB fluctuations. For the higher emissivity, the HVC signal is significantly stronger than these fluctuations and would imply a detection in our Bayesian line fit.

\subsection{Estimation of alpha and the uncertainties via cross validation}

\begin{figure}[tp]
    \includegraphics{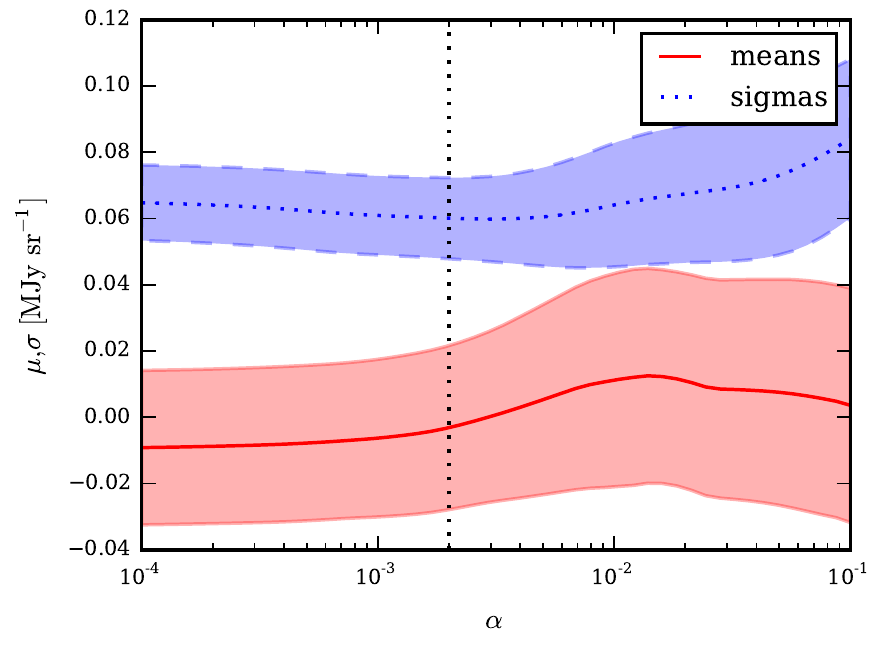}
    \caption{Results of the cross validation to investigate our choice of the regularisation strength. The solid lines correspond to the mean and standard deviation of the residual emission in the test sample. Contours correspond to $1\sigma$ uncertainties. The vertical line illustrates our choice of $\alpha$.}
    \label{fig:alpha_cv}
\end{figure}

To further validate the regularisation strength $\alpha$ in Eq. (\ref{eq:elastic_net}), we used structured cross validation \citep[e.g.][]{picard1984} with the HVC mask as kernel. For this purpose, we shifted the wide HVC mask to random positions in our field. The GLM was then computed on the data outside of this mask (training sample). Subsequently, we compared the data and the model inside the mask (test sample) by inspecting the residual mean and its standard deviation. This was done for 1000 different, random HVC mask positions. The resulting means and $1\sigma$ uncertainties of residual mean and residual standard deviation in the test sample are shown in Fig. \ref{fig:alpha_cv} for a range of regularisation strengths.

If the regularisation is too strong, the model will eventually fail and will not properly account for the complexity of the data. Based on the simulations, we chose $\alpha=2\times10^{-3}$. This is supported by the findings of the present cross validation which shows that this value is the strongest possible regularisation before the model begins to lose accuracy.

We used the same technique to estimate the uncertainties for a fixed value of the regularisation strength $\alpha=2\times 10^{-3}$ and find that the mean residual FIR intensity is $-0.003 \pm 0.025 \, \rm MJy\,sr^{-1}$. The standard deviation, equivalent to the amplitude of the CIB anisotropies for the present field-size is $0.060 \pm 0.012 \,\rm MJy\,sr^{-1}$. 

The same cross validation analysis for the standard approach yields a mean of $-0.040 \pm 0.129 \,\rm MJy\,sr^{-1}$ and a standard deviation of $0.129 \pm 0.054 \,\rm MJy\,sr^{-1}$. The larger uncertainties with respect to the GLM illustrate another advantage of the novel method. We note, however, that the standard deviation of the residuals differs significantly between the two methods. We discuss the implications further in Sect. \ref{sect:disc_glm}.

\subsection{HVC dust emissivities in simulations}

\begin{figure}[tp]
    \includegraphics{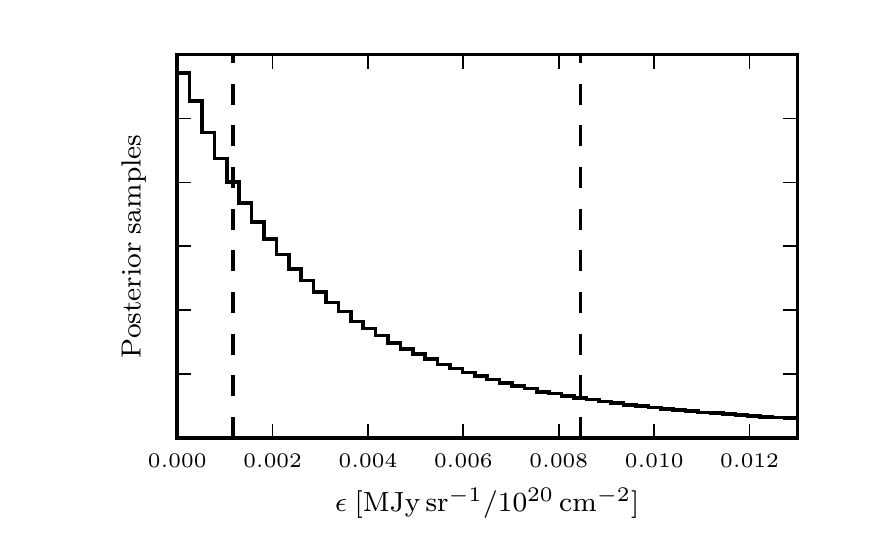}
    \caption{Stacked posterior distribution for the HVC dust emissivity $\epsilon$ from simulated data. The procedure is similar to the one described in Sect. \ref{sect:hvc_dust}. The lines in the histogram indicate the 16th and 50th percentile.}
    \label{fig:triangle_sim}
\end{figure}

To cross-check our result on the HVC dust emissivity (Sect. \ref{sect:hvc_dust}), we quantified its posterior distribution using simulated data in which no HVC dust emission is present. To generate the simulated FIR intensity maps, we used a smooth shape for the GLM coefficients (Fig. \ref{fig:simulation_input}). After removing the foreground emission, we evaluated the HVC dust emissivity using a similar procedure to the one described in Sect. \ref{sect:hvc_dust}. We present the stacked posterior distribution for all 1000 simulations in Fig. \ref{fig:triangle_sim}. The 16th, 50th, 84th and 99.87th percentiles correspond to a dust emissivity of 0.001, 0.008, 0.111, and 0.555$\,\rm MJy\,sr^{-1}/10^{20}\,\rm cm^{-2}$, respectively.

Despite the absence of a HVC signal in these simulations, the emissivities are lower than the values derived from real data (Fig. \ref{fig:triangle}). The large values for the 84th and 99.87th percentile are the consequence of the large sample of simulations and yield a heavy tail in the posterior.

\section{Discussion}
\label{ch:discussion}

\subsection{Quality of the foreground model}
\label{sect:disc_glm}

To address the potential dust content of HVC125, a reliable estimation of the foreground FIR intensity is the most challenging step and of the utmost importance.

After early studies of the \ion{H}{i}-dust correlation did not account for this foreground emission \citep{wakker1986}, the often-applied standard approach assumes a simple linear relation between FIR intensity and \ion{H}{i} column density for different \ion{H}{i} column density maps \citep[Eq. \ref{eq:classic}, e.g.][]{miville2005b, planck2011_xxiv}. This approach is limited by the uncertain separation of \ion{H}{i} in different foreground and cloud components. Furthermore, it only has a low number of degrees of freedom and hence cannot properly treat the multiple, complex features at different radial velocities. Thus, clouds with different emissivities or illumination by the ISRF cannot be accounted for if they are in the same column density map. In the case of HVC125, the separation between HVC emission and foreground emission is straightforward and the results are not affected by uncertainties introduced by different definitions of $N^{\rm HVC}$ and $N^{\rm local}$. Commonly, a clean separation of the different components for the more complex intermediate-velocity and local gas is hardly feasible. The residual histogram (Fig. \ref{fig:residual_histograms}) shows that the standard approach yields a broad and asymmetric residual signal. Moreover, Table \ref{tab:classic} shows that the fit parameters vary strongly and are poorly determined for different spatial masks.

We apply a GLM to use each individual channel of the \ion{H}{i} observations, restricted by the regularisation to overcome the degeneracy between the individual, correlated channels and to avoid overfitting. We justified the choice of the regularisation strength via simulations and cross validation, yielding consistent results. The success is well demonstrated by the narrow, symmetric distribution in the FIR intensity histogram. The results of our cross validation yields no bias with a mean residual of $0.003\pm 0.025\,\rm MJy\,sr^{-1}$. The mean standard deviation of the residual maps of $0.060\pm 0.012\,\rm MJy\,sr^{-1}$ result from the underlying CIB fluctuations \citep{planck2011_xviii, planck2014_xxx}. Moreover, the parameters are barely affected by different mask sizes, unlike the standard approach (Table \ref{tab:modern}). The spectrum of GLM coefficients illustrates that the local \ion{H}{i} does not contribute uniformly to the foreground FIR intensity, but different features need to be accounted for. Aside from the complex, bright emission from the gas around $0\,\rm km\,s^{-1}$, we find different filaments at intermediate velocities. Within the GLM framework, these features are recognised and the FIR intensity is successfully modelled.

The investigation of the standard deviation via cross validation generates significantly different results between the standard approach and the GLM. Here, the former is in agreement with studies of high-latitude fields of similar sizes using the very same approach \citep{planck2011_xviii, planck2014_xxx}. However, our simulations show that we can properly reconstruct the different simulated CIB properties with the GLM approach. Accordingly, the tension in the strength of the CIB fluctuations requires further investigation in a future study. We note at this point that we constrain ourselves to this particular field on the sky which can only provide very limited insights into the global CIB properties.

\subsection{Dust content of the HVC}

We find that the foreground-corrected FIR intensity map does not contain any significant contribution from the HVC. This is clearly seen in the posterior distribution of the emissivity $\epsilon$ (Fig. \ref{fig:triangle}, top), which strongly favours zero dust emissivity. Given our model, there is a $99.87\%$ probability that the emissivity of the HVC is below $0.02\,\rm MJy\,sr^{-1}/10^{20}\,cm^{-2}$. This is an order or magnitude lower than typical Galactic values found by \citet{planck2011_xxiv}.

Similar non-detections have been obtained in other studies of dust in HVCs \citep{planck2011_xxiv, saul2014}. A noteworthy exception is HVC complex M \citep{peek2009}. However, complex M is on the transition of HVC/IVC classification and could possibly be related to the IV arch \citep{wakker2001}.

Because of the very low kinetic temperature, its compact structure, and high brightness temperature, HVC125 is one of the most promising candidates for the detection of dust in HVCs. Moreover, the head-tail structure can indicate the formation of dust and molecules via the increased pressure \citep{gillmon2006, guillard2009, roehser2014}. Our non-detection shows that even for this candidate, the upper limit is significantly below typical ISM dust emissivities.

\section{Conclusion}
\label{ch:conclusion}

\subsection{Summary}
\label{ch:summary}

To explore the properties and the origin of HVCs, their potential dust content can be a powerful tool. We pointed out the importance and the difficulty of estimating the foreground contribution to the FIR intensity in the classical framework. Without an accurate and robust determination of this foreground emission, it is not possible to evaluate the expected, very faint dust emission from HVCs.

We have presented a new approach to address this issue by applying a GLM to evaluate the correlation of atomic neutral hydrogen and FIR dust emission. The GLM offers the opportunity to attribute an individual dust emissivity to each spectral channel of the \ion{H}{i} data. To regularise the fit, we introduced linear penalty terms for the GLM coefficients. The investigation of the residual FIR intensity shows that the GLM yields significantly lower residual emission than the standard approach. Furthermore, the distribution function of the residual is more symmetric, has fewer outliers, and the derived model is more robust to variations of the spatial area that is approximated.

After correcting for the foreground dust emission via this GLM, we analysed the potential dust content of the HVC. Using a line fit in the Bayesian framework, we derived that the dust emissivity at $857\,\rm GHz$ is $0.021\,\rm MJy\,sr^{-1}/10^{20}\,cm^{-2}$ at the 99.87\% confidence level. This is more than an order of magnitude lower than typical ISM emissivities. This shows that even for this promising candidate with low kinetic temperatures, high brightness temperature, and head-tail structure, the detection of dust is not feasible via the correlation of dust and neutral gas.

\subsection{Outlook}
\label{ch:outlook}

For the future search of dust in HVCs, we plan to apply a similar method to a larger sample of clouds and improve the signal-to-noise ratio by stacking them. Furthermore, spectroscopic observations of molecular gas tracers such as CO and OH \citep{allen2015} can help to shed light on the dust content of HVCs.

In another upcoming study, we will combine the different FIR frequencies as well as data on the gaseous content of the Milky Way such as \ion{H}{i} and CO. Combined with a proper treatment of the CIB characteristics such the spatial covariance that needs to be considered for the analysis, we are confident that we will provide full-sky information on the dust emissivities, the $X_{\rm CO}$ conversion factor between molecular hydrogen and CO intensity ($\equiv N_{\rm H_2}/W_{\rm CO}$), and the CIB.


\begin{acknowledgements}
We thank the referee for providing very helpful feedback, especially for the role of the CIB anisotropies. We thank the Deutsche Forschungsgemeinschaft (DFG) for the project funding under grants KE757/7-1 to -3. Based on observations with the 100\,m telescope of the MPIfR (Max-Planck-Institut f\"{u}r Radioastronomie) at Effelsberg. D.L. is a member of the Bonn-Cologne Graduate School of Physics and Astronomy (BCGS). L.F. is a member of the International Max Planck Research School (IMPRS) for Astronomy and Astrophysics at the Universities of Bonn and Cologne. Reproduced with permission from Astronomy \& Astrophysics, \copyright\, ESO.
\end{acknowledgements}

\bibliographystyle{aa}
\bibliography{references}

\begin{appendix}
\section{Estimators in GLM reconstruction}
\label{ch:appendix_estimator}

\begin{figure}[tp]
    \includegraphics{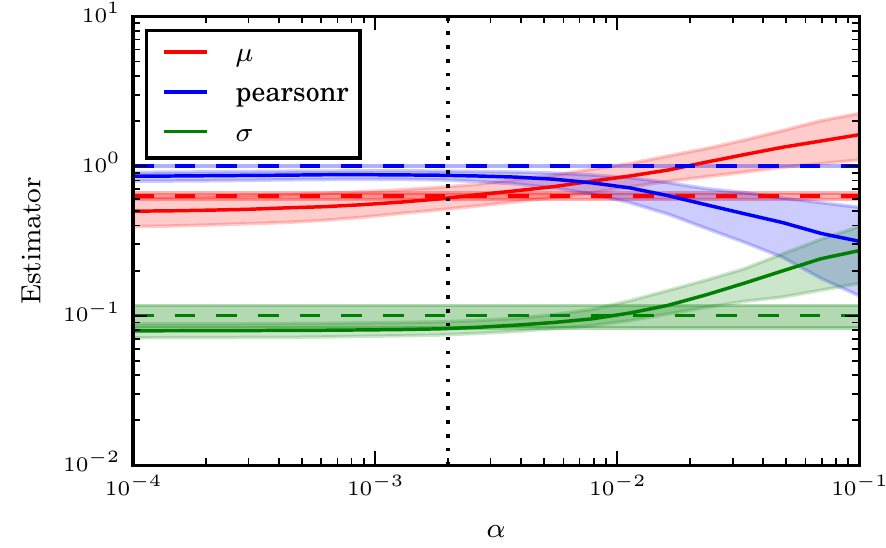}
    \caption{Evaluation of the reconstructed CIB mean $\mu$, CIB fluctuation amplitude $\sigma$ and Pearson's $r$ of input CIB image and reconstruction. This is based on the spiky input GLM coefficients (second panel from the top in Fig. \ref{fig:simulation_input}). Dashed lines indicate the input, solid lines the reconstructed quantities. Contours correspond to $1\sigma$ uncertainties.}
    \label{fig:simulation_evaluation_spiky}
\end{figure}

\begin{figure}[tp]
    \includegraphics{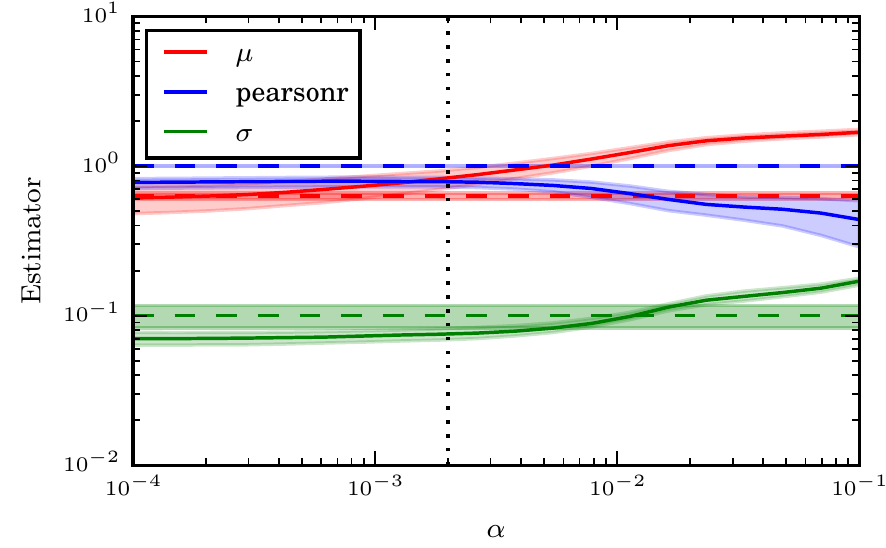}
    \caption{Same as Fig. \ref{fig:simulation_evaluation_spiky}, but for a flat spectrum of GLM coefficients (bottom panel in Fig. \ref{fig:simulation_input}).}
    \label{fig:simulation_evaluation_flat}
\end{figure}

We present the evaluation of the simulations for the spiky and flat GLM coefficients (Figs. \ref{fig:simulation_evaluation_spiky} and \ref{fig:simulation_evaluation_flat}, respectively). See Sect. \ref{sect:simulation_evaluation} for a detailed description of the results and their implications.

\end{appendix}

\end{document}